\documentclass[runningheads]{llncs}
\usepackage{graphicx}
\PassOptionsToPackage{hyphens}{url}\usepackage{hyperref}
\begin{document}
\title{NetFlow Datasets for Machine Learning-based Network Intrusion Detection Systems}
\titlerunning{NetFlow Datasets for ML-based NIDS}
\author{Mohanad Sarhan\inst{1}\and
Siamak Layeghy\inst{1}\and
Nour Moustafa\inst{2}\and
Marius Portmann\inst{1}}
\authorrunning{Sarhan et al.}
\institute{University of Queensland, Brisbane QLD 4072, Australia 
\\
\email{m.sarhan@uq.edu.au};
\email{siamak.layeghy@uq.net.au};
\email{marius@ieee.org}\\
\and
University of New South Wales, Canberra ACT 2612, Australia 
\email{nour.moustafa@unsw.edu.au}}

\maketitle              % typeset the header of the contribution
\begin{abstract}
Machine Learning (ML)-based Network Intrusion Detection Systems (NIDSs) have proven to become a reliable intelligence tool to protect networks against cyberattacks. Network data features has a great impact on the performances of ML-based NIDSs. However, evaluating ML models often are not reliable, as each ML-enabled NIDS is trained and validated using different data features that may do not contain security events. Therefore, a common ground feature set from multiple datasets is required to evaluate an ML model's detection accuracy and its ability to generalise across datasets. This paper presents NetFlow features from four benchmark NIDS datasets known as UNSW-NB15, BoT-IoT, ToN-IoT, and CSE-CIC-IDS2018 using their publicly available packet capture files.  In a real-world scenario, NetFlow features are relatively easier to extract from network traffic compared to the complex features used in the original datasets, as they are usually extracted from packet headers. The generated Netflow datasets have been labelled for solving binary- and multiclass-based learning challenges. Preliminary results indicate that NetFlow features lead to similar binary-class results and lower multi-class classification results amongst the four datasets compared to their respective original features datasets. The NetFlow datasets are named NF-UNSW-NB15, NF-BoT-IoT, NF-ToN-IoT, NF-CSE-CIC-IDS2018 and NF-UQ-NIDS are published at \cite{netflow_datasets_2020} for research purposes.

\end{abstract}
\section{Introduction}
Anomaly-based Network Intrusion Detection Systems (NIDSs) aim to learn and extract complex network data behaviours to classify incoming traffic into an attack or a benign class \cite{garcia2009anomaly}. Network attack vectors can be obtained from various features transmitted through network traffic, such as packet counts/sizes, protocols, services and flags. Each network attack's type has a different identifying pattern, known as a set of events that may compromise the security principles of networks if undetected  \cite{6779523}. These patterns are learned from network traffic data, hence the importance of data collection for Machine Learning (ML) training and evaluation stages. Real network data is challenging to obtain due to security and privacy issues. Also, production networks do not generate labelled flows, which is necessary for following a supervised ML learning method. Therefore, researchers have created synthetic datasets through virtual test-beds that are publicly available for research purposes \cite{SHIRAVI2012357}. The NIDS datasets contain labelled network flows that are made up of certain features extracted from network traffic. These features are pre-determined by the datasets' authors based on their domain knowledge and tools used during their extraction.

Over the past few years, researchers have utilised the datasets using their original set of features. However, as these feature sets are unique and often completely different than each other, researchers have been unable to evaluate their proposed learning models on multiple datasets using a specific set of features. Moreover, due to their complex techniques of extraction, these network feature sets might not be feasible for collection or storage in some high-traffic live networks. Therefore, we have converted four well-known modern NIDS datasets into NetFlow format. NetFlow is a widely deployed protocol of network traffic collection \cite{claise2004cisco}. Obtaining NetFlow features from existing different datasets will enable researchers in evaluating ML models across various datasets using the same set of features. Moreover, it will also determine the performance of NetFlow features in detecting various attack types present in the datasets. Section \ref{lm} illustrates the limitations faced by existing datasets and how they can be overcome. Section \ref{NF} explains the importance and methodology of creating NetFlow datasets as well as describing the distribution of the newly created datasets. Finally, we evaluate the new datasets in Section \ref{evaluation}, by comparing their binary-class and multi-class classification performance to the original features of their respective datasets. The contribution of this paper is providing the research community with five NetFlow datasets using four existing benchmark datasets along with an initial set of results collected while evaluating the new datasets using binary-class and multi-class classification experiments.

\section{Limitations of Existing Datasets}
\label{lm}
Due to the complexity in obtaining real-world labelled benign and attack network flows, researchers have generated benchmark NIDS datasets. They are made publicly available to be used in the training and testing stages of the proposed ML detection model. There are more than 15 available NIDS datasets in the field \cite{RING2019147}, each containing labelled network data flows. These datasets reflect real network benign behaviour combined with synthetic attack scenarios. Each dataset contains certain attack categories conducted over a test-bed network. The corresponding packets are captured in their native format packet capture (pcap) and certain network features are extracted. A key stage of designing an ML-based NIDS is the selection of features used in the classification stages. These features must be feasible in count and extraction's complexity for efficient storage and collection. In addition, they should represent sufficient and valuable information to enable the ML model for effective classification performance. Until the time of this paper's writing, there is no benchmarked or standard set of features to be used in generated NIDS datasets. Therefore, datasets' authors have utilised their domain knowledge to extract pre-identified key network features that they believe would aid in the classification process. As a result, each available dataset was created with their own unique set of network features. 

The variance of information represented in each dataset has caused limitations in the field that keeps aggravating with the new releases and production of NIDS datasets. The two main issues of having different feature sets in benchmark datasets are; 1. dimensional overload due to collection and storage of various features, some of which are irrelevant and 2. inability of evaluating an ML model's performance generalisation across multiple NIDS datasets using a targeted or a proposed feature set. We believe this may have caused a gap between the extensive academic research conducted and the actual deployments of ML-based NIDS models into the real world. Identifying the ideal set of network features to be used in NIDS datasets has been an ongoing research topic over the last decade. However, due to the subjection to the datasets used in the experiments, the identified feature sets have been custom to each dataset. These sets are also subjected to the feature selection techniques and ML models used to identify and evaluate them respectively. Moreover, due to the differences in datasets', the selected or identified features can not be evaluated using other datasets, simply due to their absence. The rest of this section discusses four of the most recent and common publicly available NIDS datasets. These datasets have been released within the last five years so they represent modern behavioural network attacks.

\begin{itemize}

\item UNSW-NB15- The Cyber Range Lab of the Australian Centre for Cyber Security (ACCS) released the widely used, UNSW-NB15, dataset in 2015. The IXIA PerfectStorm tool was utilised to generate a hybrid of real benign network activities as well as synthetic attack scenarios. Tcpdump tool was implemented to capture a total of 100 GB of pcap files. Argus and Bro-IDS now called Zeek, and twelve additional algorithms were used to extract the dataset's original 49 features \cite{moustafa-slay-2015}. The dataset contains 2,218,761 (87.35\%) benign flows and 321,283 (12.65\%) attack ones, that is, 2,540,044 flows in total.

\item BoT-IoT- The Cyber Range Lab of the Australian Centre for Cyber Security (ACCS) designed a realistic network environment that consists of normal and botnet traffic \cite{DBLP:journals/corr/abs-1811-00701}. The Ostinato and Node-red tools were utilised to generate the non-IoT and IoT traffic respectively. A total of 69.3GB of pcap files were captured and Argus tool was used to extract the dataset's original 42 features. The dataset contains 477 (0.01\%) benign flows and 3,668,045 (99.99\%) attack ones, that is, 3,668,522 flows in total.

\item ToN-IoT- A recent heterogeneous dataset released in 2020 \cite{fesz-dm97-19} that includes telemetry data of Internet of Things (IoT) services, network traffic of IoT networks and operating system logs. In this paper, we utilise the portion containing network traffic flows. The dataset is made up of a large number of attack scenarios conducted in a realistic representation of a medium-scale network at the Cyber Range Lab by ACCS. Bro-IDS, now called Zeek, was used to extract the dataset's original 44 features. The dataset is made up of 796,380 (3.56\%) benign flows and 21,542,641 (96.44\%) attack samples, that is, 22,339,021 flows in total.

\item CSE-CIC-IDS2018- A dataset released by a collaborative project between the Communications Security Establishment (CSE) \& Canadian Institute for Cybersecurity (CIC) in 2018 \cite{sharafaldin-habibi-lashkari-ghorbani-2018}. The victim network was designed in a realistic manner of five different organisational departments and an additional server room. The benign packets were generated by realistic network events using the abstract behaviour of human users. The attack scenarios were executed by one or more machines outside the target network. The CICFlowMeter-V3 tool was used to extract the original dataset's 75 features. The full dataset has 13,484,708 (83.07\%) benign flows and 2,748,235 (16.93\%) attack flows, that is, 16,232,943 flows in total.
\end{itemize}

\begin{figure}[ht]
    \centering
    \includegraphics[width=10cm, height=7cm]{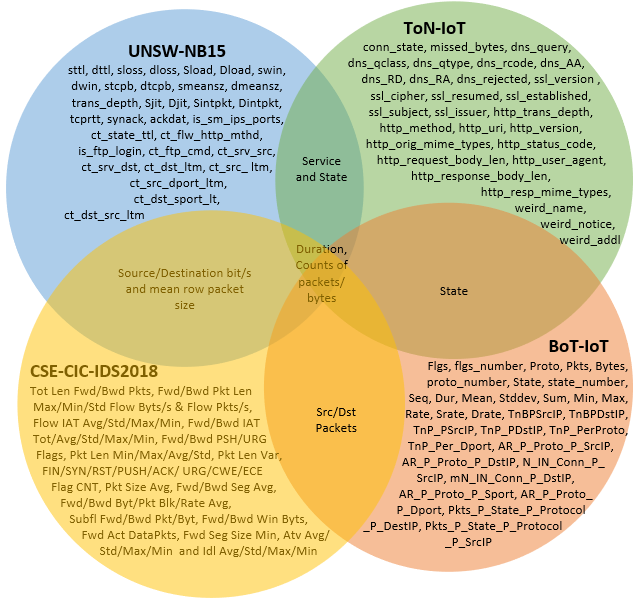}
    \caption{Venn diagram illustrating the overlap and differences of the features of four NIDS datasets discussed in this paper}
    \label{SF}
\end{figure}

The extracted features of the datasets are unique in their design, Figure \ref{SF} shows the datasets’ shared features which are only a few, making it challenging for researchers to measure the performance of their proposed models using the same set of features across the four datasets. Other differences include, UNSW-NB15 and CSE-CIC-IDS2018 datasets have more benign than attack samples whereas the ToN-IoT and BoT-IoT datasets has a significantly larger number of attack than benign ones. UNSW-NB15 and ToN-IoT have approximately the same numbers of original features but CSE-CIC-IDS2018 has almost doubled their figures and BoT-IoT having slightly lower numbers. UNSW-NB15, BoT-IoT and CSE-CIC-IDS2018 original feature sets contain handcrafted features that are not present in network traffic but are statistically measured from other features, such as the average or sum of the number of bytes transferred over the last 100 seconds. All these differences lead to the necessity of generating a common ground feature set to be shared amongst the datasets. This will enable researchers to evaluate their proposed ML-based NIDS model's performance across various network designs and attack scenarios. 

% \begin{table}[ht]\footnotesize
% \centering
% \caption{Datasets' shared features}
% \label{SF}
% \begin{tabular}{|l|l|}
% \hline
% \textbf{Dataset}             & \textbf{Features in common}                                              \\\hline
% UNSW-NB15 \& ToN-IoT         & Service and State                                              \\\hline
% UNSW-NB15 \& CSE-CIC-IDS2018 & Source/Destination bit/s and mean row packet size                        \\\hline
% ToN-IoT \& CSE-CIC-IDS2018   & N/A                                                              \\\hline
% All                          & Counts and Duration of packets/bytes \\ \hline
% \end{tabular}
% \end{table}

\section{NetFlow Datasets}
\label{NF}
\subsection{NetFlow}
Collecting and recording network traffic is necessary to monitor and analyse networks. There are two main trends for this process, capturing the complete network traffic, i.e. traffic packets, and capturing a summary of network packets in the form of flows. While packet capturing provides full access to traffic history for the network and security analysis, it is not scalable to large and medium-size networks. Indeed, even for small networks, it might necessitate large-capacity data storage to record short period of network traffic. The large volume of such dataset not only makes it difficult for the analysis, but it also faces privacy and security concerns. The alternative method is capturing network traffic summary as flows, which is more common in the networking industry due to its scalability. A network flow identifies a sequence of packets between two endpoints with some common attributes. The packets flow can be unidirectional or bidirectional. These common attributes include; source/destination IP address and L4 (transport layer) ports, and the L4 protocol. These shared attributed are often referred to as the five-tuple. The information provided by network flows are not only essential to analyse network traffic for network security, but they are also a necessity for an appropriate network planning \cite{Li2013}. 

The network flows can be represented in various formats where the NetFlow is the de-facto industry standard developed and proposed by Darren and Barry Bruins from Cisco in 1996 \cite{Kerr2001}. NetFlow is utilised in network traffic collection and analysis. Other network hardware manufacturers have also implemented and adopted their protocols such as \textit{NetStream} by Huawei, \textit{Jflow} by Juniper, \textit{Cflow} by Alcatel-Lucent, \textit{Rflow} by Ericsson and \textit{s-flow} that is supported by 3Com/HP, Dell, and Netgear. In response to the need for a universal standard of flow information, the Internet Engineering Task Force (IETF) has developed a new protocol, named Internet Protocol Flow Information Export (IPFIX) which is based on Cisco NetFlow. Similar to the NetFlow, IPFIX considers a flow to be any number of packets observed in a specific time slot and sharing some properties such as the five-tuple. While NetFlow v5 (version 5) is unidirectional (ingress), the later versions of NetFlow such as version 9 are bidirectional and include much more fields \cite{Li2013}. The actual number of fields that is possible to export using NetFlow v9 is much larger including source and destination Autonomous System (AS) numbers, Border Gateway Protocol (BGP) next-hop address, IPv6 fields, Multiprotocol Label Switching (MPLS) fields, etc.  \cite{CiscoSystems2011}. 

\begin{figure}[ht]
    \centering
    \includegraphics[width=7cm, height=3cm]{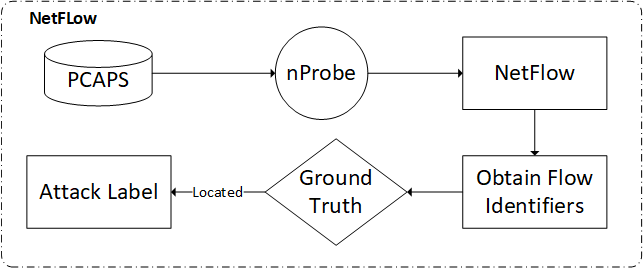}
    \caption{NetFlow datasets' extraction and labelling}
    \label{fig:Netflow}
\end{figure}

Some probe makers such as ntop~\cite{Ntopng2017} have made it possible to even include the application layer (L7) protocols in the exported flow information. NetFlow makes it possible to convert any available dataset into a common ground feature set. Accomplishing that, researchers would be able to compare datasets efficiently and most importantly evaluate their proposed ML-based NIDS models using the same set of features across various datasets and attack types. Most of the real world's network devices such as routers and switches are capable of extracting NetFlow records hence the motivation of evaluating the performance of NetFlow features in terms of attack detection. NetFlow defines a flow as a unidirectional transmission of packets sharing the five-tuple. Therefore, the number of samples in NetFlow datasets are less compared to the original datasets. There are multiple versions of NetFlow, version 9 is one of the most commonly used as it is compatible with most of the recent network devices and include additional features. Figure \ref{fig:Netflow} illustrates the procedure of converting and labelling the original datasets' pcaps into NetFlow-based format.

\subsection{Conversion}

We utilised the publicly available pcap files of each dataset to generate the NetFlow datasets, nProbe tool by Ntop \cite{Ntopng2017} was utilised to convert the pcaps into NetFlow version 9 format and selecting 12 features to be extracted. Table \ref{nf} lists the extracted NetFlow features along with their brief description. Using nProbe we create a text file listing the pcaps path of the original datasets, we specify NetFlow version 9 due to its popularity. The dump format is chosen as text flows each feature separated by a comma (,) to be utilised as CSV files. The maximum number of flows in a file is 100m dumped in a maximum of 100m seconds, and nProbe is set not to modify the original pcaps timestamps. Finally, we create two label features by matching the five flow identifiers; source/destination IPs and ports and protocol to the ground truth attack events published with the original datasets. If a flow is located in the attack events it would be labelled as an attack, class 1, in the binary label and its respective attack's type would be recorded in the attack label, otherwise, the record is labelled as a benign flow, class 0. Table \ref{ds} compares the properties of the datasets, in both their original and NetFlow format, in terms of Feature Extraction (FE) tool, the number of features, files size and benign to attack samples ratio. The full nProbe command used to extract the NetFlow features is;\begin{verbatim}
nprobe /c --pcap-file-list 'Source pcaps list' -V 9 -n none -T
%OUT_BYTES%OUT_PKTS%L4_DST_PORT%IPV4_DST_ADDR%IPV4_SRC_ADDR
%PROTOCOL%L4_SRC_PORT%IN_BYTES%IN_PKTS%L7_PROTO%TCP_FLAGS
%FLOW_DURATION_MILLISECONDS --dump-path 'Destination' --dump-format 
t --csv-separator , --max-log-lines 100000000 
--dont-reforge-timestamps --dump-frequency 100000000
\end{verbatim} 

\begin{table}[ht]\footnotesize
\centering
\caption{Extracted NetFlow features}
\label{nf}
\begin{tabular}{|l|l|}
\hline
\multicolumn{1}{|c|}{\textbf{Feature}} & \multicolumn{1}{c|}{\textbf{Description}} \\ \hline
IPV4\_SRC\_ADDR                        & IPv4 source address                       \\ \hline
IPV4\_DST\_ADDR                        & IPv4 destination address                  \\ \hline
L4\_SRC\_PORT                          & IPv4 source port number                   \\ \hline
L4\_DST\_PORT                          & IPv4 destination port number              \\ \hline
PROTOCOL                               & IP protocol identifier byte               \\ \hline
TCP\_FLAGS                             & Cumulative of all TCP flags               \\ \hline
L7\_PROTO                              & Layer 7 protocol (numeric)                \\ \hline
IN\_BYTES                              & Incoming number of bytes                  \\ \hline
OUT\_BYTES                             & Outgoing number of bytes                  \\ \hline
IN\_PKTS                               & Incoming number of packets                \\ \hline
OUT\_PKTS                              & Outgoing number of packets                \\ \hline
FLOW\_DURATION\_MILLISECONDS           & Flow duration in milliseconds             \\ \hline
\end{tabular}
\end{table}

\begin{itemize}

\item NF-UNSW-NB15- The NetFlow-based format of the UNSW-NB15 dataset, named NF-UNSW-NB15, has been developed and labelled with its respective attack categories. The total number of data flows are 1,623,118 out of which 72,406 (4.46\%) are attack samples and 1,550,712 (95.54\%) are benign. The attack samples are further classified into nine subcategories, Table \ref{un} represents the NF-UNSW-NB15 dataset's distribution of all flows.

\begin{table}[ht]\footnotesize\centering
\caption{NF-UNSW-NB15 distribution}
\label{un}
\resizebox{\textwidth}{!}{%
\begin{tabular}{|l|l|l|}
\hline
\multicolumn{1}{|c|}{\textbf{Class}} & \multicolumn{1}{c|}{\textbf{Count}} & \multicolumn{1}{c|}{\textbf{Description}}                                                     \\ \hline
Benign                               & 1550712                             & Normal unmalicious flows                                                                      \\ \hline
Fuzzers &
  19463 &
  \begin{tabular}[c]{@{}l@{}}An attack in which the attacker sends large amounts of random data which cause a system \\ to crash and also aim to discover security vulnerabilities in a system.\end{tabular} \\ \hline
Analysis                             & 1995                                &  \begin{tabular}[c]{@{}l@{}} A group that presents a variety of threats that target web applications through ports, \\emails and scripts. \end{tabular} \\ \hline
Backdoor                             & 1782                                & \begin{tabular}[c]{@{}l@{}} A technique that aims to bypass security mechanisms by replying to specific constructed \\client applications.   \end{tabular}\\ \hline
DoS &
  5051 &
  \begin{tabular}[c]{@{}l@{}}Denial of Service is an attempt to overload a computer system’s resources with the aim \\ of preventing access to or availability of its data.\end{tabular} \\ \hline
Exploits                             & 24736                               & \begin{tabular}[c]{@{}l@{}}Are sequences of commands controlling the behaviour of a host through a known \\vulnerability. \end{tabular} \\ \hline
Generic                              & 5570                                & A method that targets cryptography and causes a collision with each block-cipher.          \\ \hline
Reconnaissance                       & 12291                               & A technique for gathering information about a network host and is also known as a probe.   \\ \hline
Shellcode                            & 1365                                & A malware that penetrates a code to control a victim’s host.                                 \\ \hline
Worms                                & 153                                 & Attacks that replicate themselves and spread to other computers.                          \\ \hline
\end{tabular}
}
\end{table}

\item NF-BoT-IoT- An IoT NetFlow-based dataset generated using the BoT-IoT dataset, named NF-BoT-IoT. The features were extracted from the publicly available pcap files and the flows were labelled with their respective attack categories. The total number of data flows are 600,100 out of which 586,241 (97.69\%) are attack samples and 13,859 (2.31\%) are benign. There are four attack categories in the dataset, Table \ref{bo} represents the NF-BoT-IoT distribution of all flows.

\begin{table}[ht]\footnotesize\centering
\caption{NF-BoT-IoT distribution}
\label{bo}
\resizebox{\textwidth}{!}{%
\begin{tabular}{|l|l|l|}
\hline
\multicolumn{1}{|c|}{\textbf{Class}} & \multicolumn{1}{c|}{\textbf{Count}} & \multicolumn{1}{c|}{\textbf{Description}}                                                     \\ \hline
Benign                               & 13859                             & Normal unmalicious flows                                                                      \\ \hline
Reconnaissance &
  470655 &
  A technique for gathering information about a network host and is also known as a probe. \\ \hline
DDoS                             & 56844                                &  \begin{tabular}[c]{@{}l@{}}Distributed Denial of Service is an attempt similar to DoS but has multiple \\ different distributed sources.\end{tabular} \\ \hline
DoS                             & 56833                                &  \begin{tabular}[c]{@{}l@{}}An attempt to overload a computer system’s resources with the aim of preventing access\\ to or availability of its data.\end{tabular}\\ \hline
Theft &
  1909 &
  \begin{tabular}[c]{@{}l@{}}A group of attacks that aims to obtain sensitive data such as data theft and keylogging\end{tabular} \\ \hline
\end{tabular}
}
\end{table}

\item NF-ToN-IoT- We utilised the publicly available pcaps of the ToN-IoT dataset to generate its NetFlow records, leading to a NetFlow-based IoT network dataset called NF-ToN-IoT. The total number of data flows are 1,379,274 out of which 1,108,995 (80.4\%) are attack samples and 270,279 (19.6\%) are benign ones, Table \ref{ton} lists and defines the distribution of the NF-ToN-IoT dataset.

\begin{table}[ht]\footnotesize\centering
\caption{NF-ToN-IoT distribution}
\label{ton}
\resizebox{\textwidth}{!}{%
\begin{tabular}{|l|l|l|}
\hline
\multicolumn{1}{|c|}{\textbf{Class}} &
  \multicolumn{1}{c|}{\textbf{Count}} &
  \multicolumn{1}{c|}{\textbf{Description}} \\ \hline
Benign &
  270279 &
  Normal unmalicious flows \\ \hline
Backdoor &
  17247 &
  \begin{tabular}[c]{@{}l@{}}A technique that aims to attack remote-access computers by replying to specific constructed \\ client applications \end{tabular} \\ \hline
DoS &
  17717 &
  \begin{tabular}[c]{@{}l@{}}An attempt to overload a computer system’s resources with the aim of preventing access to or\\ availability of its data.\end{tabular} \\ \hline
DDoS &
  326345 &
  \begin{tabular}[c]{@{}l@{}}An attempt similar to DoS but has multiple \\ different distributed sources.\end{tabular} \\ \hline
Injection &
  468539 &
  \begin{tabular}[c]{@{}l@{}}A variety of attacks that supply untrusted inputs that aim to alter the course of \\ execution, with SQL and Code injections two of the main ones.\end{tabular} \\ \hline
MITM &
  1295 &
  \begin{tabular}[c]{@{}l@{}} Man In The Middle is a method that places an attacker between a victim and host with which \\ the victim is trying to communicate, with the aim of intercepting traffic and communications.\end{tabular} \\ \hline
Password &
  156299 &
  covers a variety of attacks aimed at retrieving passwords by either brute force or sniffing. \\ \hline
Ransomware &
  142 &
  \begin{tabular}[c]{@{}l@{}} An attack that encrypts the files stored on a host and asks for compensation in exchange for \\ the decryption technique/key.\end{tabular} \\ \hline
Scanning &
  21467 &
  \begin{tabular}[c]{@{}l@{}}A group that consists of a variety of techniques that aim to discover information about networks \\ and hosts, and is also known as probing.\end{tabular} \\ \hline
XSS &
  99944 &
  \begin{tabular}[c]{@{}l@{}}Cross-site Scripting is a type of injection in which an attacker uses web applications to send \\ malicious scripts to end-users.\end{tabular} \\ \hline
\end{tabular}
}
\end{table}

\item NF-CSE-CIC-IDS2018- We utilised the original pcap files of the CSE-CIC-IDS2018 dataset to generate a NetFlow-based dataset called NF-CSE-CIC-IDS2018. The total number of flows are 8,392,401 out of which 1,019,203 (12.14\%) are attack samples and 7,373,198 (87.86\%) are benign ones, Table \ref{cse} represents the dataset's distribution.

\begin{table}[ht]\footnotesize\centering
\caption{NF-CSE-CIC-IDS2018 distribution}
\label{cse}
\resizebox{\textwidth}{!}{%
\begin{tabular}{|l|l|l|}
\hline
\multicolumn{1}{|c|}{\textbf{Class}} &
  \multicolumn{1}{c|}{\textbf{Count}} &
  \multicolumn{1}{c|}{\textbf{Description}} \\ \hline
Benign &
  7373198 &
  Normal unmalicious flows \\ \hline
BruteForce &
  287597 &
   \begin{tabular}[c]{@{}l@{}}A technique that aims to obtain usernames and password credentials by accessing a list of \\predefined possibilities \end{tabular}\\ \hline
Bot &
  15683 &
  \begin{tabular}[c]{@{}l@{}}An attack that enables an attacker to remotely control several hijacked computers to perform\\ malicious activities.\end{tabular} \\ \hline
DoS &
  269361 &
  \begin{tabular}[c]{@{}l@{}}An attempt to overload a computer system’s resources with the aim of preventing access to or \\ availability of its data.\end{tabular} \\ \hline
DDoS &
  380096 &
  An attempt similar to DoS but has multiple different distributed sources. \\ \hline
Infiltration &
  62072 &
  \begin{tabular}[c]{@{}l@{}}An inside attack that sends a malicious file via an email to exploit an application and is \\followed by a backdoor that scans the network for other vulnerabilities\end{tabular} \\ \hline
Web Attacks &
  4394 &
  A group that includes SQL injections, command injections and unrestricted file uploads \\ \hline
\end{tabular}
}
\end{table}

\item NF-UQ-NIDS- A comprehensive dataset, merging all the aforementioned datasets. The newly published dataset represents the benefits of shared dataset feature sets, where the merging of multiple smaller ones is possible. This will eventually lead to a bigger and more universal NIDS datasets containing flows from multiple network setups and different attack settings. An additional label feature identifying the original dataset of each flow. This can be used to compare the same attack scenarios conducted over two or more different test-bed networks. The attack categories have been modified to combine all parent categories. Attacks named DoS attacks-Hulk, DoS attacks-SlowHTTPTest, DoS attacks-GoldenEye and DoS attacks-Slowloris have been renamed to the parent DoS category. Attacks named DDOS attack-LOIC-UDP, DDOS attack-HOIC and DDoS attacks-LOIC-HTTP have been renamed to DDoS. Attacks named FTP-BruteForce, SSH-Bruteforce, Brute Force -Web and Brute Force -XSS have been combined as a brute-force category. Finally, SQL Injection attacks have been included in the injection attacks category. The NF-UQ-NIDS dataset has a total of 11,994,893 records, out of which 9,208,048 (76.77\%) are benign flows and 2,786,845 (23.23\%) are attacks. Table \ref{uq} lists the distribution of the final attack categories.

\begin{table}[ht]\scriptsize
\centering
\caption{NF-UQ-NIDS distrubution}
\label{uq}
\begin{tabular}{|l|l|}
\hline
\textbf{Class} & \textbf{Count} \\ \hline
Benign         & 9208048        \\ \hline
DDoS           & 763285         \\ \hline
Reconnaissance & 482946          \\ \hline
Injection      & 468575         \\ \hline
DoS            & 348962         \\ \hline
Brute Force    & 291955         \\ \hline
Password       & 156299         \\ \hline
XSS            & 99944          \\ \hline
Infilteration  & 62072          \\ \hline
Exploits       & 24736          \\ \hline
Scanning       & 21467          \\ \hline
Fuzzers        & 19463          \\ \hline
Backdoor       & 19029          \\ \hline
Bot            & 15683          \\ \hline
Generic        & 5570           \\ \hline
Analysis       & 1995           \\ \hline
Theft      & 1909           \\ \hline
Shellcode      & 1365           \\ \hline
MITM           & 1295           \\ \hline
Worms          & 153            \\ \hline
Ransomware     & 142            \\ \hline
\end{tabular}
\end{table}

% \usepackage{graphicx}
\begin{table}[ht]\footnotesize
\centering
\caption{Datasets' comparison}
\label{ds}
\resizebox{\textwidth}{!}{%
\begin{tabular}{|l|l|l|l|l|l|}
\hline
\multicolumn{1}{|c|}{\textbf{Dataset}} &
  \multicolumn{1}{c|}{\textbf{\begin{tabular}[c]{@{}c@{}}Release \\ year\end{tabular}}} &
  \multicolumn{1}{c|}{\textbf{Feature extraction tool}} &
  \multicolumn{1}{c|}{\textbf{\begin{tabular}[c]{@{}c@{}}Number \\ of features\end{tabular}}} &
  \multicolumn{1}{c|}{\textbf{\begin{tabular}[c]{@{}c@{}}CSV size\\ (GB)\end{tabular}}} &
  \multicolumn{1}{c|}{\textbf{\begin{tabular}[c]{@{}c@{}}Benign to attack \\ samples ratio\end{tabular}}} \\ \hline
UNSW-NB15          & 2015 & Argus, Bro-IDS and MS SQL & 49 & 0.55 & 8.7 to 1.3 \\
NF-UNSW-NB15       & 2020 & nProbe                    & 12 & 0.11 & 9.6 to 0.4 \\ \hline
BoT-IoT             & 2018 & Argus                   & 42 & 0.95 & 0 to 10 \\
NF-BoT-IoT          & 2020 & nProbe                    & 12 & 0.05 & 0.2 to 9.8 \\ \hline
ToN-IoT            & 2020 & Bro-IDS                   & 44 & 3.02 & 0.4 to 9.6 \\
NF-ToN-IoT         & 2020 & nProbe                    & 12 & 0.09 & 2.0 to 8.0 \\ \hline
CSE-CIC-IDS2018    & 2018 & CICFlowMeter-V3           & 75 & 6.41 & 8.3 to 1.7 \\
NF-CSE-CIC-IDS2018 & 2020 & nProbe                    & 12 & 0.58 & 8.8 to 1.2 \\ \hline
NF-UQ-NIDS         & 2020 & nProbe                    & 12 & 1.0 & 7.7 to 2.3 \\ \hline
\end{tabular}%
}
\end{table}

\end{itemize}

\section{Evaluation}
\label{evaluation}
The detection performance of an ML classifier is evaluated using the newly published NetFlow datasets as a use case and compared to the original datasets. We drop the flow identifiers such as IDs, source/destination IP and ports, timestamps and start/end time to avoid bias towards attacking or victim nodes. For UNSW-NB15, we additionally drop Time To Live (TTL) based features i.e., sttl, dttl and ct\_state\_ttl, due to their extreme correlation with the labels. Furthermore, we utilise the min-max normalisation technique to scale all datasets' values between 0 to 1. Finally, we apply an Extra Trees ensemble classifier, made up of 50 randomised decision trees estimators. The chosen classifier belongs to the 'trees' family and has proven to achieve reliable performances on NIDS datasets. Due to the extreme imbalance in all datasets' binary-class and multi-class labels, we set a custom class weight parameter, using Equation \ref{e1}. To reliably evaluate the datasets, we conduct five cross-validation splits and collect the average metrics such as accuracy, Area Under the Curve (AUC), F1 Score, Detection Rate (DR), False Alarm Rate (FAR) and time required in microseconds (\textmu s) to predict a single test sample.

\begin{equation}Weight_{class}=\frac{Total Samples Count}{Number Of Classes \times Class Samples Count}\label{e1}\end{equation}

\subsection{Binary-class Classification}
In this experiment, we evaluate the attack detection performance of the NetFlow datasets compared to the original datasets. Table \ref{m1} lists the accuracy, AUC, F1 score, DR, FAR and prediction time results. The NF-UNSW-NB15 dataset achieved slightly lower performance than the UNSW-NB15 dataset, with almost the same DR but higher FAR, however, it used less time to predict the samples. The overall accuracy achieved by the NF-UNSW-NB15 dataset is 98.62\% compared to 99.25\% when using the UNSW-NB15 dataset. The NF-BoT-IoT dataset has achieved slightly lower classification performance, i.e. 93.70\% DR and 0.97 F1 Score, compared to its parent BoT-IoT dataset which achieved a 100\% DR and 1.00 F1 Score. The almost perfect results achieved bu the BoT-IoT has been deemed unreliable in a recent study \cite{alothman2020stateoftheart}, due to its extreme class imbalance of attack and benign samples which is unrealistic in a real-world network. The NF-ToN-IoT dataset's performance was superior to its original ToN-IoT dataset, achieving a 99.67\% DR and 0.37\% FAR, it also consumed less prediction time. The accuracy achieved is 99.66\% proving its significance compared to the ToN-IoT dataset, 97.86\%. The NF-CSE-CIC-IDS2018 dataset performance was less efficient than the CSE-CIC-IDS2018 dataset achieving a similar DR of 94.71\% but a higher FAR of 4.59\%, however significantly less time was consumed in prediction. The overall accuracy achieved is 95.33\%, significantly lowering the 98.31\% accuracy of the CSE-CIC-IDS2018 dataset. The merged NF-UQ-NIDS dataset achieved an accuracy of 97.25\%, a DR of 95.66\% and a FAR of 2.27\%, achieving a reliable classification performance of 20 different attack categories.

\begin{table}[ht]\scriptsize
\centering
\caption{Binary-class classification results}
\label{m1}
\resizebox{\textwidth}{!}{%
\begin{tabular}{|l|r|r|r|r|r|r|}
\hline
\textbf{Dataset} &
  \multicolumn{1}{c|}{\textbf{Accuracy}} &
  \multicolumn{1}{c|}{\textbf{AUC}} &
  \multicolumn{1}{c|}{\textbf{F1 Score}} &
  \multicolumn{1}{c|}{\textbf{DR}} &
  \multicolumn{1}{c|}{\textbf{FAR}} &
  \multicolumn{1}{c|}{\textbf{Prediction Time (\textmu s)}} \\ \hline
UNSW-NB15         & 99.25\% & 0.9545 & 0.92 & 91.25\% & 0.35\% & 10.05 \\
NF-UNSW-NB15       & 98.62\% & 0.9485 & 0.85 & 90.70\% & 1.01\% & 7.79  \\ \hline
BoT-IoT         & 100.00\% & 0.9948 & 1.00 & 100.00\% & 1.05\% & 7.62 \\
NF-BoT-IoT       & 93.82\% & 0.9628 & 0.97 & 93.70\% & 1.13\% & 5.37  \\ \hline
ToN-IoT            & 97.86\% & 0.9788 & 0.99 & 97.86\% & 2.10\% & 8.93  \\
NF-ToN-IoT         & 99.66\% & 0.9965 & 1.00 & 99.67\% & 0.37\% & 6.05  \\ \hline
CSE-CIC-IDS2018    & 98.31\% & 0.9684 & 0.94 & 94.75\% & 1.07\% & 23.01 \\
NF-CSE-CIC-IDS2018 & 95.33\% & 0.9506 & 0.83 & 94.71\% & 4.59\% & 17.04 \\ \hline
NF-UQ-NIDS        & 97.25\% & 0.9669 & 0.94 & 95.66\% & 2.27\% & 14.35 \\ \hline
\end{tabular}%
}
\end{table}

\begin{figure}[ht]
    \centering
    \includegraphics[width=8cm, height=4cm]{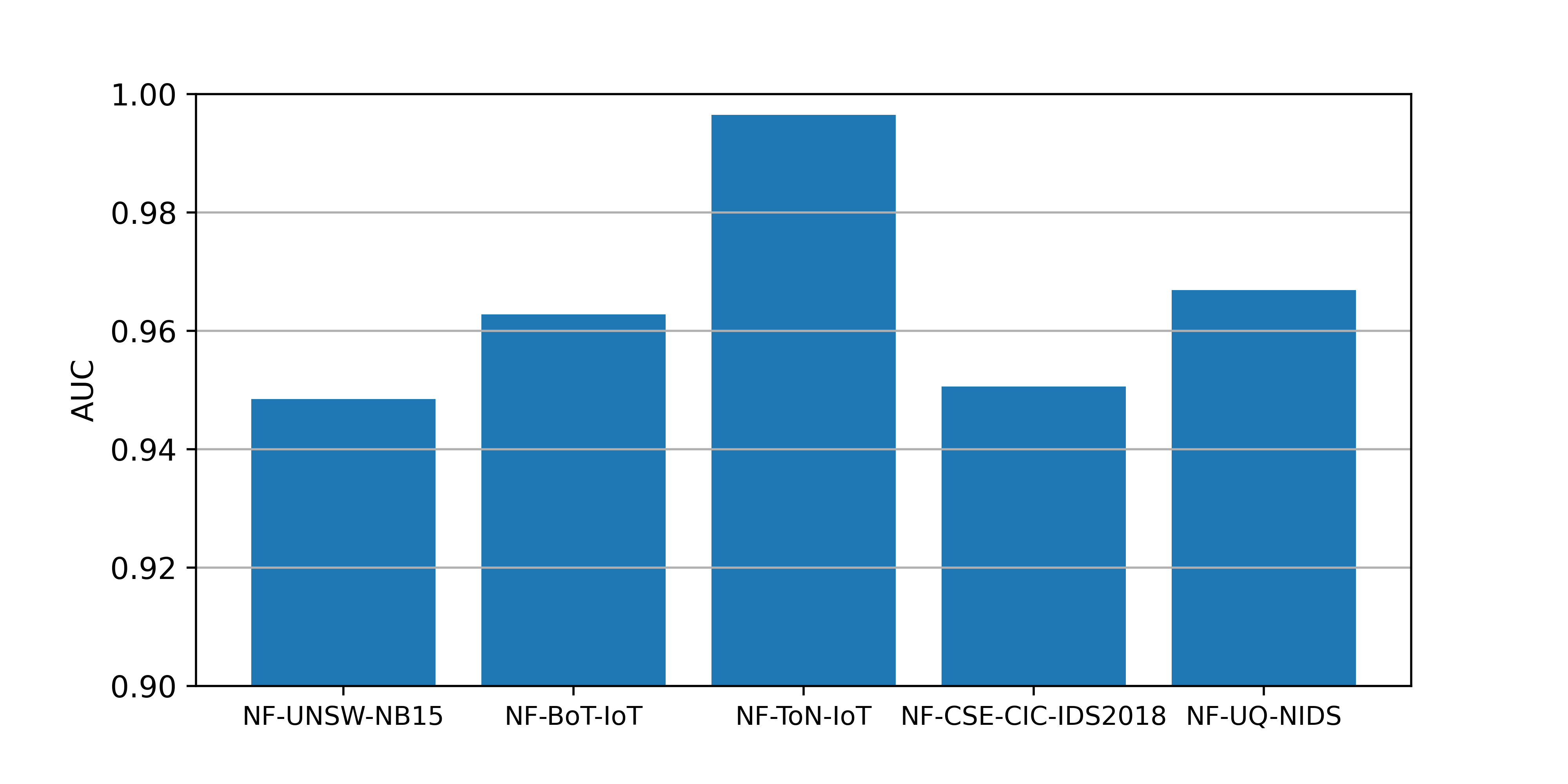}
    \caption{Binary-class classification's AUC}
    \label{we}
\end{figure}

Figure \ref{we} displays the AUC achieved using the Extra Trees classifier on the four newly published NetFlow-based datasets. This comparison is conducted by using the same set of features across all datasets. This fair comparison demonstrates the benefit of the newly published datasets, which was not possible to achieve due to each dataset's unique set of features. Overall, the NetFlow datasets containing only eight features used in the classification experiments achieved a very similar attack detection performance compared to the original 36 features of the BoT-IoT, 38 features of the UNSW-NB15 and ToN-IoT datasets and the 77 features of the CSE-CIC-IDS2018 dataset. We noticed a consistent prediction time decrease in using all the NetFlow datasets. Therefore, in terms of feasibility and practicality in real-world networks, using NetFlow features might lead to an overall superior performance if additional metrics are measured such as storage and computation power required to extract and store the utilised features.

\subsection{Multi-class Classification}
In this experiment, we measure the DR and F1 score of each attack's type present in each dataset. Tables \ref{tab:my-table1}, \ref{bot} \ref{tab:my-table}, \ref{tab:my-table3} and \ref{tab:uq} list each attack's type DR and F1 score of the NF-UNSW-NB15, NF-BoT-IoT, NF-ToN-IoT, NF-CSE-CIC-IDS2018 and NF-UQ-NIDS datasets respectively. The average accuracy and prediction time are calculated and the results are compared to their respective original datasets. In Table \ref{tab:my-table1}, we can conclude that by using the NF-UNSW-NB15 dataset, we can increase the DR of analysis, backdoor, DoS, fuzzers, shellcode and worms attacks, however, it was inefficient against generic attacks. The overall accuracy achieved which is 97.62\% is slightly lower than the UNSW-NB15 dataset, 98.19\%, due to the number of miss-correctly classified samples, however, the prediction time consumed was slightly lower. 
 
 \begin{table}[ht]\scriptsize
\centering
\caption{NF-UNSW-NB15 multi-class classification results}
\label{tab:my-table1}
\begin{tabular}{l|r|r|r|r|}
\cline{2-5}
                                               & \multicolumn{2}{c|}{\textbf{UNSW-NB15}} & \multicolumn{2}{c|}{\textbf{NF-UNSW-NB15}} \\ \hline
\multicolumn{1}{|c|}{\textbf{Class Name}} &
  \multicolumn{1}{c|}{\textbf{DR}} &
  \multicolumn{1}{c|}{\textbf{F1 Score}} &
  \multicolumn{1}{c|}{\textbf{DR}} &
  \multicolumn{1}{c|}{\textbf{F1 Score}} \\ \hline
\multicolumn{1}{|l|}{{Benign}}          & 99.72\%              & 1.00             & 99.02\%               & 0.99               \\ \hline
\multicolumn{1}{|l|}{{Analysis}}        & 4.39\%               & 0.03             & 28.28\%               & 0.15               \\ \hline
\multicolumn{1}{|l|}{{Backdoor}}        & 13.96\%              & 0.08             & 39.17\%               & 0.17               \\ \hline
\multicolumn{1}{|l|}{{DoS}}             & 13.63\%              & 0.18             & 31.84\%               & 0.41               \\ \hline
\multicolumn{1}{|l|}{{Exploits}}        & 83.25\%              & 0.80             & 81.04\%               & 0.82               \\ \hline
\multicolumn{1}{|l|}{{Fuzzers}}         & 50.50\%              & 0.57             & 62.63\%               & 0.55               \\ \hline
\multicolumn{1}{|l|}{{Generic}}         & 86.08\%              & 0.91             & 57.13\%               & 0.66               \\ \hline
\multicolumn{1}{|l|}{{Reconnaissance}}  & 75.90\%              & 0.80             & 76.89\%               & 0.82               \\ \hline
\multicolumn{1}{|l|}{{Shellcode}}       & 53.61\%              & 0.59             & 87.91\%               & 0.75               \\ \hline
\multicolumn{1}{|l|}{{Worms}}           & 5.26\%               & 0.09             & 52.91\%               & 0.55               \\ \hline
\multicolumn{1}{|l|}{\textbf{Weighted Average}}         & \textbf{98.19\%}              & \textbf{0.98}             & \textbf{97.62\%}               & \textbf{0.98}               \\ \hline
% \multicolumn{1}{|l|}{\textbf{Accuracy}}        & \multicolumn{2}{c|}{\textbf{98.19\%}}   & \multicolumn{2}{c|}{\textbf{97.62\%}}      \\ \hline
\multicolumn{1}{|l|}{\textbf{Prediction Time (\textmu s)}} & \multicolumn{2}{c|}{\textbf{9.94}}      & \multicolumn{2}{c|}{\textbf{9.35}}         \\ \hline
\end{tabular}
\end{table}

Table \ref{bot} shows that the BoT-IoT dataset is achieving almost perfect multi-classification performances of a 100\% accuracy and 1 F1 Score. Again, these results might be unreliable due to the extreme imbalance mentioned in \cite{alothman2020stateoftheart}. In addition, there might be certain 'hidden label' features, such as the TTL-based features in the UNSW-NB15 dataset, that are extremely correlated to the attack types present in the dataset. The NF-BoT-IoT dataset was unreliable in the detection of the DDoS and DoS attacks. However, it achieved a 90\% DR against reconnaissance and theft attacks. Although it achieved a lower DR of 73.58\% and F1 Score of 0.77, the NetFlow dataset maintained the lower prediction time compared to the BoT-IoT dataset.

% \usepackage{graphicx}
\begin{table}[ht]\scriptsize
\centering
\caption{NF-BoT-IoT multi-class classification results}
\label{bot}
\begin{tabular}{l|r|r|r|r|}
\cline{2-5}
\textbf{}                                     & \multicolumn{2}{c|}{\textbf{BoT-IoT}} & \multicolumn{2}{c|}{\textbf{NF-BoT-IoT}} \\ \hline
\multicolumn{1}{|c|}{\textbf{Class Name}} &
  \multicolumn{1}{c|}{\textbf{DR}} &
  \multicolumn{1}{c|}{\textbf{F1 Score}} &
  \multicolumn{1}{c|}{\textbf{DR}} &
  \multicolumn{1}{c|}{\textbf{F1 Score}} \\ \hline
\multicolumn{1}{|l|}{{Benign}}         & 99.58\%             & 0.99            & 98.65\%              & 0.43              \\ \hline
\multicolumn{1}{|l|}{{DDoS}}           & 100.00\%            & 1.00            & 30.37\%              & 0.28              \\ \hline
\multicolumn{1}{|l|}{{DoS}}            & 100.00\%            & 1.00            & 36.33\%              & 0.31              \\ \hline
\multicolumn{1}{|l|}{{Reconnaissance}} & 100.00\%            & 1.00            & 89.95\%              & 0.90              \\ \hline
\multicolumn{1}{|l|}{{Theft}}          & 91.16\%             & 95.37           & 88.06\%              & 0.18              \\ \hline
\multicolumn{1}{|l|}{\textbf{Weighted Average}} &
  \textbf{100.00\%} &
  \textbf{1.00} &
  \textbf{73.58\%} &
  \textbf{0.77} \\ \hline
\multicolumn{1}{|l|}{\textbf{Prediction Time (\textmu s)}} &
  \multicolumn{2}{c|}{\textbf{12.63}} &
  \multicolumn{2}{c|}{\textbf{9.19}} \\ \hline
\end{tabular}%
\end{table}

In Table \ref{tab:my-table}, the NF-ToN-IoT dataset increased the DR of DoS attacks but lowered the DDoS, injection, MITM, password, scanning and XSS attacks compared to the ToN-IoT dataset. Further analysis is required to identify which features of the original dataset were critical in the detection of the missed attacks and to be added to the NetFlow dataset. Overall, in multi-class classification, the NF-ToN-IoT dataset was not as effective in terms of overall accuracy and prediction time compared to the ToN-IoT dataset. It achieved a low prediction accuracy of 56.34\% and a high prediction time of 21.21 \textmu s. However, a binary-class classification deemed it was very efficient, therefore, it seems like the ML classifier is detecting the overall pattern of attacks present in the dataset, but not the pattern of individual attacks. We suspect that specific features present in the original dataset contain payload information that was enabling the ML classifier to detect certain attack types. Further analysis is required to investigate which features from the ToN-IoT dataset are necessary to identify each attack's type.

\begin{table}[ht]\scriptsize
\centering
\caption{NF-ToN-IoT multi-class classification results}
\label{tab:my-table}
\begin{tabular}{l|l|l|l|l|}
\cline{2-5}
\textbf{}                                      & \multicolumn{2}{c|}{\textbf{ToN-IoT}}       & \multicolumn{2}{c|}{\textbf{NF-ToN-IoT}}    \\ \hline
\multicolumn{1}{|l|}{\textbf{Class Name}}      & \textbf{DR} & \textbf{F1 Score} & \textbf{DR} & \textbf{F1 Score} \\ \hline
\multicolumn{1}{|l|}{{Benign}}     & 89.97\% & 0.94 & 98.97\% & 0.99 \\ \hline
\multicolumn{1}{|l|}{{Backdoor}}   & 98.05\% & 0.31 & 99.22\% & 0.98 \\ \hline
\multicolumn{1}{|l|}{{DDoS}}       & 96.90\% & 0.98 & 63.22\% & 0.72 \\ \hline
\multicolumn{1}{|l|}{{DoS}}        & 53.89\% & 0.57 & 95.91\% & 0.48 \\ \hline
\multicolumn{1}{|l|}{{Injection}}  & 96.67\% & 0.96 & 41.47\% & 0.51 \\ \hline
\multicolumn{1}{|l|}{{MITM}}       & 66.25\% & 0.16 & 52.81\% & 0.38 \\ \hline
\multicolumn{1}{|l|}{{Password}}   & 86.99\% & 0.92 & 27.36\% & 0.24 \\ \hline
\multicolumn{1}{|l|}{{Ransomware}} & 89.87\% & 0.11 & 87.33\% & 0.83 \\ \hline
\multicolumn{1}{|l|}{{Scanning}}   & 75.05\% & 0.85 & 31.30\% & 0.08 \\ \hline
\multicolumn{1}{|l|}{{XSS}}        & 98.83\% & 0.99 & 24.49\% & 0.19 \\ \hline
\multicolumn{1}{|l|}{\textbf{Weighted Average}}    & \textbf{84.61\%} & \textbf{0.87} & \textbf{56.34\%} & \textbf{0.60} \\ \hline
% \multicolumn{1}{|l|}{\textbf{Accuracy}}        & \multicolumn{2}{c|}{\textbf{84.61\%}}       & \multicolumn{2}{c|}{\textbf{56.34\%}}       \\ \hline
\multicolumn{1}{|l|}{\textbf{Prediction Time (\textmu s)}} & \multicolumn{2}{c|}{\textbf{12.02}}         & \multicolumn{2}{c|}{\textbf{21.21}}         \\ \hline
\end{tabular}
\end{table}

In Table \ref{tab:my-table3}, the performance of the NF-CSE-CIC-IDS2018 dataset can prove that attacks such as FTP-bruteforce and infiltration were better detected using the NetFlow features compared to the CSE-CIC-IDS2018 features. However, Brute Force -Web, Brute Force -XSS, DDOS attack-HOIC and SQL injection attack samples were mostly undetected by using the NetFlow features. The DoS attacks-SlowHTTPTest attack samples were fully undetected by the ML classifier. Similar to the NF-ToN-IoT dataset, the ML classifier was unable to efficiently detect the pattern of certain attack types. Overall, the accuracy and prediction time achieved while using the NF-CSE-CIC-IDS2018 dataset being 71.92\% and 17.29 \textmu s respectively were lower compared to the CSE-CIC-IDS2018 dataset.

% \usepackage{graphicx}
\begin{table}[ht]\scriptsize
\centering
\caption{NF-CSE-CIC-IDS2018 multi-class classification results}
\label{tab:my-table3}
\begin{tabular}{l|l|l|l|l|}
\cline{2-5}
\textbf{}                                      & \multicolumn{2}{c|}{\textbf{CSE-CIC-IDS2018}} & \multicolumn{2}{c|}{\textbf{NF-CSE-CIC-IDS2018}} \\ \hline
\multicolumn{1}{|l|}{\textbf{Class Name}}      & \textbf{DR}  & \textbf{F1 Score}  & \textbf{DR}    & \textbf{F1 Score}   \\ \hline
\multicolumn{1}{|l|}{{Benign}}                   & 89.50\%  & 0.94 & 69.83\%  & 0.82 \\ \hline
\multicolumn{1}{|l|}{{Bot}}                      & 99.92\%  & 0.99 & 100.00\% & 1.00 \\ \hline
\multicolumn{1}{|l|}{{Brute Force -Web}}         & 71.36\%  & 0.01 & 50.21\%  & 0.52 \\ \hline
\multicolumn{1}{|l|}{{Brute Force -XSS}}         & 72.17\%  & 0.72 & 49.16\%  & 0.39 \\ \hline
\multicolumn{1}{|l|}{{DDOS attack-HOIC}}         & 100.00\% & 1.00 & 45.66\%  & 0.39 \\ \hline
\multicolumn{1}{|l|}{{DDOS attack-LOIC-UDP}}     & 83.59\%  & 0.82 & 80.98\%  & 0.82 \\ \hline
\multicolumn{1}{|l|}{{DDoS attacks-LOIC-HTTP}}   & 99.93\%  & 1.00 & 99.93\%  & 0.71 \\ \hline
\multicolumn{1}{|l|}{{DoS attacks-GoldenEye}}    & 99.97\%  & 1.00 & 99.32\%  & 0.98 \\ \hline
\multicolumn{1}{|l|}{{DoS attacks-Hulk}}         & 100.00\% & 1.00 & 99.65\%  & 0.99 \\ \hline
\multicolumn{1}{|l|}{{DoS attacks-SlowHTTPTest}} & 69.80\%  & 0.60 & 0.00\%   & 0.00 \\ \hline
\multicolumn{1}{|l|}{{DoS attacks-Slowloris}}    & 99.44\%  & 0.62 & 99.95\%  & 1.00 \\ \hline
\multicolumn{1}{|l|}{{FTP-BruteForce}}           & 68.76\%  & 0.75 & 100.00\% & 0.79 \\ \hline
\multicolumn{1}{|l|}{{Infilteration}}            & 36.15\%  & 0.08 & 62.66\%  & 0.04 \\ \hline
\multicolumn{1}{|l|}{{SQL Injection}}            & 49.34\%  & 0.30 & 25.00\%  & 0.22 \\ \hline
\multicolumn{1}{|l|}{{SSH-Bruteforce}}           & 99.99\%  & 1.00 & 99.93\%  & 1.00 \\ \hline
\multicolumn{1}{|l|}{\textbf{Weighted Average}}                  & \textbf{90.28\%}  & \textbf{0.94} & \textbf{71.92\%}  & \textbf{0.80} \\ \hline
% \multicolumn{1}{|l|}{\textbf{Accuracy}}        & \multicolumn{2}{c|}{\textbf{90.28\%}}         & \multicolumn{2}{c|}{\textbf{71.92\%}}            \\ \hline
\multicolumn{1}{|l|}{\textbf{Prediction Time (\textmu s)}} & \multicolumn{2}{c|}{\textbf{24.17}}           & \multicolumn{2}{c|}{\textbf{17.29}}              \\ \hline
\end{tabular}%
\end{table}

Table \ref{tab:uq} displays the full attack identification results of the merged dataset named NF-UQ-NIDS. The chosen ML classifier was efficient in the detection of certain attack's types such as backdoor, bot, bruteforce, exploits, shellcode, DDoS and ransomware. However, attacks such as analysis, DoS, fuzzers, generic, infiltration, worms, injection, MITM, password, scanning and XSS were not reliably detected. Further analysis is required to identify the features that are critical in identifying these attacks and to add them to the NetFlow features. The overall accuracy of 70.81\% and prediction time 14.74 (\textmu s) were achieved.

\begin{table}[ht]\scriptsize
\centering
\caption{NF-UQ-NIDS multi-class classification results}
\label{tab:uq}
\begin{tabular}{l|r|r|}
\cline{2-3}
                                              & \multicolumn{2}{c|}{\textbf{NF-UQ-NIDS}} \\ \hline
\multicolumn{1}{|c|}{\textbf{Class Name}}      & \multicolumn{1}{c|}{\textbf{Detection Rate}} & \multicolumn{1}{c|}{\textbf{F1 Score}} \\ \hline
\multicolumn{1}{|l|}{{Analysis}}       & 69.63\%               & 0.21             \\ \hline
\multicolumn{1}{|l|}{{Backdoor}}       & 90.95\%               & 0.92             \\ \hline
\multicolumn{1}{|l|}{{Benign}}         & 71.70\%               & 0.83             \\ \hline
\multicolumn{1}{|l|}{{Bot}}            & 100.00\%              & 1.00             \\ \hline
\multicolumn{1}{|l|}{{Brute Force}}    & 99.94\%               & 0.85             \\ \hline
\multicolumn{1}{|l|}{{DoS}}            & 55.54\%               & 0.62             \\ \hline
\multicolumn{1}{|l|}{{Exploits}}       & 80.65\%               & 0.81             \\ \hline
\multicolumn{1}{|l|}{{Fuzzers}}        & 63.24\%               & 0.54             \\ \hline
\multicolumn{1}{|l|}{{Generic}}        & 58.90\%               & 0.61             \\ \hline
\multicolumn{1}{|l|}{{Infilteration}}  & 60.57\%               & 0.03             \\ \hline
\multicolumn{1}{|l|}{{Reconnaissance}} & 88.96\%               & 0.88            \\ \hline
\multicolumn{1}{|l|}{{Shellcode}}      & 83.89\%               & 0.15             \\
\hline
\multicolumn{1}{|l|}{{Theft}}          & 87.22\%               & 0.15             \\\hline
\multicolumn{1}{|l|}{{Worms}}          & 52.97\%               & 0.46             \\ \hline
\multicolumn{1}{|l|}{{DDoS}}           & 77.08\%               & 0.69             \\ \hline
\multicolumn{1}{|l|}{{Injection}}      & 40.58\%               & 0.50             \\ \hline
\multicolumn{1}{|l|}{{MITM}}           & 57.99\%               & 0.10             \\ \hline
\multicolumn{1}{|l|}{{Password}}       & 30.79\%               & 0.27             \\ \hline
\multicolumn{1}{|l|}{{Ransomware}}     & 90.85\%               & 0.85             \\ \hline
\multicolumn{1}{|l|}{{Scanning}}       & 39.67\%               & 0.08             \\ \hline
\multicolumn{1}{|l|}{{XSS}}            & 30.80\%               & 0.21             \\ \hline
\multicolumn{1}{|l|}{\textbf{Weighted Average}}        & \textbf{70.81\%}               & \textbf{0.79}             \\ \hline
% \multicolumn{1}{|l|}{\textbf{Accuracy}}        & \multicolumn{2}{c|}{\textbf{71.82\%}}                                                 \\ \hline
\multicolumn{1}{|l|}{\textbf{Prediction Time (\textmu s)}} & \multicolumn{2}{c|}{\textbf{14.74}}                                                   \\ \hline
\end{tabular}
\end{table}

\section{Conclusion}
This paper provides the research community with four new NIDS datasets using NetFlow features. These datasets are to be used in ML-based NIDS training and evaluation stages. The datasets are showing positive results by achieving similar binary-class detection performance compared to the complete set of their respective original datasets. However, the NF-ToN-IoT and NF-CSE-CIC-IDS2018 datasets were inefficient when conducting multi-class detection experiments. Further feature analysis is required to identify the strength of each NetFlow feature, and how these datasets can be improved by adding key features from the original datasets to aid in the detection of missed attack types. While further experiments are required, the published NetFlow datasets offer a promising performance. The datasets serve two advantages; 1. the level of complexity and resources required to collect and store NetFlow features are lower and 2. proposed ML models can be evaluated using the same set of features across various datasets' attack types. Overall, the practicality and initial performance of NetFlow features' collection and attack detection, requires increased attention and interest by researchers in applying them into the real-world for ML-based NIDS. Future works include enhancing the current datasets with additional NetFlow features which can potentially improve both the binary and multi-class classification performances. Finally, key features from the original datasets required to detect certain attack types must be identified to be included in NetFlow features.

\bibliography{access.bib}
% \printbibliography
%






% LNCS Homepage, \url{http://www.springer.com/lncs}. Last accessed 4

% \end{thebibliography}
\end{document}